\documentstyle[version2,preprint,aps]{revtex}
\begin{document}
\draft
\begin{title}

Coulomb drag between disordered two-dimensional electron gas
layers

\end{title}
\author{Lian Zheng and A.H. MacDonald}
\begin{instit}

Department of Physics,
Indiana University, Bloomington, IN 47405

\end{instit}

\begin{abstract}

We derive and evaluate expressions for the frictional
Coulomb drag between disordered
two-dimensional electron gas layers.
Our derivation is based on the
memory-function formalism and the expression for the
drag reduces to previously known results in the ballistic limit.
We find that Coulomb drag is appreciably enhanced by disorder
at low temperatures when the mean-free-path
within a layer is comparable to or shorter than
 the layer separation.
In high mobility two-dimensional electron gas systems,
where the drag has been studied experimentally,
the effect of disorder on the drag is negligible at attainable
temperatures.  We predict that an enhancement due to disorder and a
crossover in the temperature-dependence and layer-separation
dependence will be observable at low temperatures
in moderate and low mobility samples.

\end{abstract}
\narrowtext

\section{Introduction}

Double-layer two dimensional electron gas (2DEG) systems,
where electrons are confined to nearby
parallel planes,  are expected to exhibit many novel phenomena
due to interlayer electron-electron
interaction.  For example, in strong magnetic fields
Coulomb effects are
expected to produce new incompressible ground states
that exhibit the fractional quantum Hall effect\cite{ahm1} and
to cause the collapse of certain integer quantum Hall
effect gaps\cite{ahm2}.  In zero magnetic field, it has
been suggested that Wigner crystallization in double layer
systems is favored by interlayer Coulomb interactions\cite{neilson}
and that in the case of electron-hole systems, excitonic
superfluity could result\cite{spfl} from interlayer interactions.
In recent experiments by
Gramila et al. on electron-electron double layer
systems\cite{gramila1,gramila2} and in
similar experiment by Sivan et al. on electron-hole systems\cite{e-h},
the strength of interlayer interactions
was studied directly by measuring the frictional drag of
one two-dimensional electron gas layer on another.
In these experiments a current flowing in one layer tends to
induce a current in nearby layers.
If no current is allowed to flow in the nearby layer
an electric field develops whose influence cancels the
frictional force between the layers.  The transresistance, defined as
the ratio of the induced voltage in the second layer to
the applied current in the first layer, directly measures
the rate at which momentum is transferred from the current
carrying 2DEG to its neighbor.

Drag between spatially separated electron systems
due to Coulomb interactions between carriers was first considered by
Pogrebinskii\cite{sov} and Price\cite{price}.
The experiments of Gramila et al.\cite{gramila1,gramila2} and of
Sivan et al.\cite{e-h} have stimulated recent theoretical
attention, especially to the case of drag between two-dimensional
electron layers at low
temperatures\cite{gramila1,gramila2,solomon,jauho,mahan1,rojo,takis}.
The interlayer Coulomb drag is caused by fluctuations in
the density of electrons in each layer since 2D layers with
charge uniformly distributed will not exert any frictional
forces upon each other.
In this paper we examine for the
first time the possibility of enhanced frictional drag
between disordered layers due to the diffusive nature of
long-wavelength long-time electron density fluctuations.
Disorder is known\cite{aa,plee} to enhance interaction effects
and to lead to violations of Fermi liquid theory for
individual two-dimensional electron gas layers.
In the diffusive regime,
where wavelengths are longer than the mean free path of
the electrons and times are longer than the electron scattering time,
the electron density-density response function possesses a
diffusion pole\cite{forster} in momentum-frequency space.
In perturbation theory the diffusion pole arises from
dressing the electron-electron vertex\cite{plee}
with corrections arising from impurity potential scattering.
At shorter distances or shorter times the disorder vertex
corrections are not important.
We find that disorder enhances the interlayer drag
at low temperatures, changing the temperature dependence
of the drag from $T^2$ to $-T^2 \ln T$.
In very high-mobility samples (on which existing
experiments have been performed) or in samples with small
layer separations the corrections due to disorder become
important only at extremely low temperatures.
For samples with lower mobility or more widely separated
layers the influence on the interlayer scattering rate
from disorder scattering should be easily measurable.

Most\cite{gramila1,solomon,jauho} previous work on interlayer
friction has been based on Boltzmann transport theory which
cannot capture disorder enhanced interaction effects.
(An exception is the work of Vasilopoulos and co-workers\cite{takis}.)
In Section II we present a derivation of the expression
for the frictional transresistance based on the memory
function formalism\cite{forster} which is sufficiently
general to treat the case of disordered 2DEG layers.
Although phonon-mediated\cite{gramila2,takis,myphmed}
interlayer interactions are not discussed explicitly
in this paper the expression we derive in Section II
are sufficiently general that other coupling mechanisms
can be incorporated as an effective-interlayer interaction
potential.  Similarly particle-particle interaction
vertex corrections, which are
probably quantitatively important especially
for electron-hole double layer systems\cite{e-h},
can also be incorporated into the results we derive in
Section II as a contribution to the effective
electron-electron interaction\cite{we}.  Readers interested only
in the application to disordered double-layer
systems should proceed to Section III where we
discuss how disorder within the layers influences the interlayer
friction.  The interlayer scattering rate of typical
electron-electron double layer samples is evaluated numerically
in Section IV.  We show that the temperatures
below which disorder becomes important decreases very
rapidly with increasing mobility.  A
brief summary of our findings concludes the paper in
Section V.

\section{Memory Function Formalism Derivation}

The memory function formalism provide a very convenient method
for deriving a flexible expression for the transresistance of double
layer 2DEG systems.
The Kubo current-current correlation function formula
for the conductivity is
first converted into a force-force
correlation function expression for the resistivity
by making use of Mori's\cite{mori} projection operator.
This force-force correlation function is then evaluated at
lowest order in the screened interlayer interaction to
obtain an approximate expression for the transresistance.
The advantage of using the force-force correlation function
rather than the current-current correlation function
is that it yields a reasonable approximation even when
evaluated at lowest order\cite{mf1,mf2}.  At this level
the results are physically equivalent to
momentum-balance\cite{takis} approximations or to
relaxation-time approximations in a Boltzmann-transport
approach\cite{mahan2}.  The derivation for the situation
of present interest is sketched in the following paragraphs.

For notational simplicity we restrict ourself to the case of
zero magnetic field so that the
current is in the same direction as the
applied electric field and the conductance,
therefore, forms a 2$\times$2
matrix with respect to the layer indices.
(The final expression for the transresistance
is equally valid\cite{magfield}
in the presence of a magnetic field.)
We need to consider only the long wavelength limit of the conductance.
To use the memory function formalism it is convenient to
write the Kubo formula in the form
\begin{equation}
\sigma_{ij}(\omega) = {\beta\over\nu} \int_{0}^{\infty}dt
e^{i\omega t} ( {\hat J}_i(t), {\hat J}_j )
\label{ku}
\end{equation}
where $\beta=1/k_B T$, $\nu$ is the
cross section area of the 2DEG layers, and
the indices $i$ and $j$ are layer labels.
$\sigma_{ij}(\omega)$ gives the current density induced in
layer $i$ due to an electric field in layer $j$.
$\hat J$ is the zero wavevector Fourier component of
the current density operator.
The inner product appearing in Eq.~(\ref{ku}) is defined by
\begin{eqnarray}
C_{AB}(t) &\equiv & ({\hat A(t)},{\hat B}) \nonumber \\
&\equiv & \beta^{-1}\int_{0}^{\beta}d\lambda\langle {\hat A}^\dagger(t),
{\hat B}(i\hbar\lambda)\rangle
\label{ip}
\end{eqnarray}
In Eq.~(\ref{ip}) the angle brackets denote thermal averages.
The following relationship can be used to change Eq.~(\ref{ku}) into
the more familiar form of the Kubo formula\cite{mf2}:
\begin{equation}
i\beta\partial_tC_{AB}(t) = {1\over\hbar}\langle[{\hat A}(t),{\hat
B}]\rangle.
\label{dtc}
\end{equation}

The projection operator method is now
used to obtain an expression for the matrix inverse
in layer indices of $C_{J_iJ_j}$.
We define a superoperator ${\cal P}$ which `projects' an operator
$\hat O$ onto the current density, and its complement ${\cal Q}$, by
\begin{eqnarray}
{\cal P}\hat O
&\equiv & \sum_k{ {\hat J}_k ({\hat J}_k,\hat O) \over
 ({\hat J}_k,{\hat J}_k ) }
\\
{\cal P} \hat O & \equiv & \hat O-{\cal Q} \hat O
\label{po}
\end{eqnarray}
It is useful to define a matrix $\chi_{ij}$:
\begin{equation}
\chi_{ij} = {\beta\over\nu} C_{J_iJ_j}(0) = ({\hat J}_i,{\hat J}_j)
= {n_ie^2\over m}\delta_{ij}.
\label{ch}
\end{equation}
Here $n_i$ is the areal density of 2D electrons in the ${i}$-th layer.
Following the usual development of the memory function
formalism\cite{forster}, we obtain an equation for the matrix
inverse of the Fourier transform of $C_{J_iJ_j}(t)$:
\begin{equation}
[C_{J_iJ_j}(z)]^{-1} = {\beta\over\nu}\chi^{-1}[-iz{\bf 1}+M(z)]
\label{em}
\end{equation}
where
\begin{equation}
M_{ij}(z) = {\beta\over\nu} (\dot{J}_i,{i\over z-{\cal Q}{\cal L}}
\dot{J}_j)\chi_{jj}^{-1}.
\label{si}
\end{equation}
A dot over an operators denotes its time derivative at $t=0$
and the Liouville superoperator is defined by ${\cal L}
{\hat O}   = [\hat H,\hat O]$, where $\hat H$ is the Hamiltonian.
In obtaining Eq.~(\ref{em}),
the time reversal invariance condition ${\cal P}\dot{J}=0$
has been applied. Combining Eq.~(\ref{ku}), Eq.~(\ref{em})
and Eq.~(\ref{si}) we obtain an expression for the resistivity
matrix:
\begin{eqnarray}
\rho_{ij}(z) &=& \chi_{ii}^{-1} M _{ij}(z)\nonumber \\
&=& \chi_{ii}^{-1}\chi_{jj}^{-1}{\beta\over\nu}
\int_{0}^{\infty}dt e^{izt}(\dot{J}_i,e^{-i{\cal Q}Lt}\dot{J}_j)
\label{ro}.
\end{eqnarray}
$\rho_{ij}$ relates the electric field in layer $i$ to the
current density in layer $j$.
With the relation $\dot{J}_i = -e/m F_i$, we obtain the force-force
correlation function expression for the trans-resistance,
\begin{equation}
\rho_{LR} = {\beta\over n_Ln_Re^2\nu}\int_{0}^{\infty}dte^{izt}
(\hat F_L,e^{-i{\cal Q}{\cal L}t}\hat F_R).
\label{rolr}
\end{equation}

It is easy to demonstrate that $\rho_{LR}$ is identically zero
in the absence of interlayer coupling, since the forces in
left and right layers are uncorrelated.
The leading contribution to the force operator
from interlayer interactions
can be expressed in terms of the
interlayer interaction potential $U_e(q)$ and the electron
density $\varrho(q)$
\begin{equation}
{\bf F}_{R(L)} = \pm {i\over\nu}\sum_{\vec q}{\vec  q}\varrho_L(\vec q)
\varrho_R(-\vec q) U_e(q).
\label{fe}
\end{equation}
To leading order in interlayer interactions
$ e^{-i{\cal Q}{\cal L}t}\hat F_i$ in Eq.~(\ref{rolr}) can be
replaced by $e^{-i{\cal L}t}\hat F_i$.
This replacement leads to the desired
force-force correlation function expression for the trans-resistance
\begin{equation}
\rho_{LR}(z) = {\beta\over n_Ln_Re^2\nu}
\int_{0}^{\infty}dt e^{izt} (F_L(t),F_R)_0.
\label{ffr}
\end{equation}
The subscript on the inner product in Eq.~(\ref{ffr})
indicates that it should be evaluated in the
absence of interlayer interactions.  Substituting the
explicit expression for the interlayer forces gives
\begin{equation}
\rho_{LR}(z) = {\beta\over2n_Ln_Re^2\nu^3}\sum_{\vec q}q^2
\int_{0}^{\infty}e^{izt}dt|U_e(q)|^2({\hat A(t)},
{\hat A}^{\dagger}(0))_0
\label{r1}
\end{equation}
with
\begin{equation}
{\hat A(t)} = \varrho_L(-\vec q,t)\varrho_R(\vec q,t).
\label{as}
\end{equation}

The correlation function for decoupled layers appearing
in Eq.~(\ref{r1}) is related to the isolated layer density
fluctuations.  Using a representation of
exact eigenstates that in the $z=0$ limit it is easy to show that
\begin{equation}
\rho_{LR} = { \pi \beta \over 2 n_L n_R e^2 }
\int {d^2 \vec q \over (2 \pi)^2 } q^2 |U_e(q)|^2
\int_{-\infty}^{\infty} d \omega
S_L(\vec q, \omega) S_R( - \vec q, - \omega)
\label{ahm1}
\end{equation}
where $S_i(\vec q,\omega) $ is the dynamic structure factor
for layer $i$:
\begin{equation}
S_i(\vec q, \omega) \equiv { 1 \over \nu }
\sum_{n,m} \exp (- \beta E_n) |\langle n | \rho_i(\vec q) | m \rangle |^2
\delta (\omega - (E_m-E_n)/\hbar )
\label{ahm2}
\end{equation}
It is usually more convenient to express the resistance in
terms of individual layer response functions rather than the dynamic
structure factor.
We relate $S_i(q,\omega)$
to the retarded density-density response function for layer
$i$, $\chi_{i}(\vec q, \omega) $ by
applying the fluctuation-dissipation theorem\cite{forster}
\begin{equation}
S_i(q,\omega) = {\hbar\over1-e^{-\hbar\omega\beta}}
{\rm Im} \chi_i(\vec q,\omega).
\label{fdearly}
\end{equation}
This gives us the final form of our expression for the drag
resistivity, which is summarized
diagrammatically in Fig.~(\ref{fa}).
\begin{equation}
\rho_{LR} = {\hbar^2\beta\over\pi n_Ln_Re^2}{1\over\nu}
\sum_{\bf q}q^2|U_e(q)|^2\int_{0}^{\infty}d\omega
{{\rm Im}\chi_{R}(q,\omega)
{\rm Im}\chi_{L}(q,\omega)\over
e^{\beta\hbar\omega}+e^{-\beta\hbar\omega}-2}.
\label{r3}
\end{equation}

\section{Disorder and Screening}

The electron density-density response
function $\chi(q,\omega)$ in Eq.~(\ref{r3})
can be obtained by applying many-body perturbation
theory methods to a 2DEG whose Hamiltonian
contains disorder and (or) interaction terms.
For a non-interacting disorder-free 2DEG the
response function can be evaluated analytically\cite{2degrf}.
Disorder leads to an enhancement in ${\rm Im}\chi(q,\omega)$
at low frequencies and long wavelengths and, as we discuss
below, can enhance the interlayer friction.  The enhancement
reflects the increased spatial correlation of states with
nearby energies in disordered systems.   The effect of
disorder on the friction can be described without
making a specific model of disorder by invoking the
Einstein relation between the conductivity and the diffusive
density-density response at long wavelengths and
low frequencies\cite{arbitrary}.
We introduce a phenomenological intralayer electron
(transport) scattering time $\tau$.
$\tau$ is related to the mobility by $\mu=e\tau/m$ and
at low temperatures is related to the mean-free-path
by $l  = \tau\hbar k_f/m$.
For $q l > 1$ or $  \omega \tau > 1$ we assume that disorder is
unimportant and approximate the density-density response
function by the non-interacting electron result\cite{2degrf}.
At zero temperature and zero disorder
\begin{eqnarray}
&&\chi_i(q,\omega) \equiv
\chi_{i}^{B}(q,\omega)\nonumber \\
&=& {dn\over d\mu}{m\over q^2}\{{q^2\over m} -
C_+|(k_fq/m)^2-(\omega+\varepsilon_q)^2/\hbar^4|^{1/2}
-C_-|(k_fq/m)^2-(\omega-\varepsilon_q)^2/\hbar^4|^{1/2}\}
\label{xb}
\end{eqnarray}
\noindent where $\varepsilon_q=\hbar q^2/2m$,
 $\mu$ is the chemical potential and the $C_{\pm}$ are:
 \begin{eqnarray}
 C_{\pm} &=& {\rm sign}(\varepsilon_q\pm\omega)\ \ \ \ \ \ {\rm if} \ \ \
 (k_fq/m)^2-(\omega\pm\varepsilon)^2/\hbar^2<0,
 \nonumber\\
 C_{\pm} &=&\pm i\ \ \  \ \ \ \ \ \ \ \ \ \ \ \ \ \ \ {\rm if}\ \ \
 (k_fq/m)^2-(\omega\pm\varepsilon_q)^2/\hbar^2.
 >0\nonumber
 \end{eqnarray}
However for $q l < 1$ and $\omega \tau < 1$ disorder becomes
important.  In this diffusive regime the electron
density-density response function is completely characterized
by the diffusion constant\cite{plee}:
\begin{eqnarray}
\chi_i(q,\omega) &\equiv &
\chi_{i}^{D}(q,\omega) \nonumber \\
&=& {dn\over d\mu}{Dq^2\over Dq^2-i\omega}
\label{xd}
\end{eqnarray}
\vskip 1ex
\centerline{ $\ \ {q<{1/l}}$\ \ \ \ \  \ \ \ \
$\omega<{1 / \tau}$}
\vskip 1ex
\noindent where $D=l^2/ 2 \tau$ is the diffusion
constant.

The derivation of the expression for the transresistivity in
the previous section is
valid up to second order in the interlayer interaction.
For the system of physical interest the interlayer interaction
is Coulombic and it is essential to include screening in
order to get qualitatively correct results.  In this
paper we adopt the usual expediency of employing
the second-order expression with the interlayer interaction
replaced by a screened interlayer interaction and argue that
this includes the most important higher-order effects.
In the disorder free limit our expression for the transresistivity
then becomes identical to those derived using other approaches
in earlier work\cite{gramila1,solomon,jauho,mahan1,takis}.
The random-phase-approximation (RPA) screened interlayer
interaction is
\begin{equation}
U_e(q,\omega) = {V_e(q)\over[1+V_a(q)\chi_L
(q,\omega)][1+V_a(q)\chi_R(q,\omega)]-V_e^2(q)\chi_L(q,\omega)
\chi_R(q,\omega)}
\label{rpa}
\end{equation}
where the bare intra- and inter-layer electron-electron interaction
potentials are $V_a(q)=2\pi e^2/q$ and $V_e(q)=V_a(q)e^{-qd}$
where $d$ is the separation between the layers.
In the above expression either the ballistic or the diffusive
form for $\chi_{L(R)}$ should be used as appropriate.  Note
that the interlayer interaction is cut-off by the factor
$e^{-qd}$ for $q > 1/d$.   The layer separation dependence
of the friction in both diffusive and ballistic limits
results from this cutoff.  Physically the cutoff reflects the
fact that charge fluctuations in one layer with a wavelength shorter
than the layer separation get averaged out when viewed from
the other layer.

The expression for the trans-resistance of Eq.~(\ref{r3})
can be split into contributions from the ballistic and diffusive
regimes. With $\rho_{LR}^{-1} \equiv n_Re^2\tau_{LR}/m$, we have
\begin{equation}
\tau_{RL}^{-1} = \tau_B^{-1} + \tau_{\Delta}^{-1}
\label{rlbd}
\end{equation}
where
\begin{equation}
\tau_B^{-1} = {\hbar^2\beta\over2\pi^2mn_L}\int_{0}^{\infty}
dqq^3\int_{0}^{\infty}d\omega |U_e|^2
{{\rm Im}[\chi_L^B(q,\omega)] {\rm Im}[\chi_R^B(q,\omega)]
\over e^{\beta\hbar\omega}+e^{-\beta\hbar\omega}-2}
\label{tbe}
\end{equation}
and
\begin{equation}
\tau_{\Delta}^{-1} =\tau_D^{-1} -
{\hbar^2\beta\over2\pi^2mn_L}\int_{0}^{1\over l}
dqq^3\int_{0}^{1\over\tau}d\omega |U_e|^2
{{\rm Im}[\chi_L^B(q,\omega)] {\rm Im}[\chi_R^B(q,\omega)]
\over e^{\beta\hbar\omega}+e^{-\beta\hbar\omega}-2}
\label{tdel}
\end{equation}
with
\begin{equation}
\tau_D^{-1} =
{\hbar^2\beta\over2\pi^2mn_L}\int_{0}^{1\over l}
dqq^3\int_{0}^{1\over\tau}d\omega |U_e|^2
{{\rm Im}[\chi_L^D(q,\omega)] {\rm Im}[\chi_R^D(q,\omega)]
\over e^{\beta\hbar\omega}+e^{-\beta\hbar\omega}-2}
\label{tde}
\end{equation}
$\tau_B^{-1}$ is the result for
a disorder-free 2DEGs and
$\tau_{\Delta}^{-1}$ is the correction due to the
the enhanced fluctuations at long wavelengths and low frequencies
in disordered systems.
In the next section we discus the evaluation of these expressions.

\section{Numerical Results and Discussion}

The dependence of the interlayer scattering rate on temperature
and on layer separation depends on whether
the interlayer scattering is dominated by $\tau_B^{-1}$ or
$\tau_D^{-1}$.
In Fig.~(\ref{fb}) and Fig.~(\ref{fc})
we show numerical results for $\tau_B$ and
$\tau_{\Delta}$ as a functions of temperature calculated
for two different values of layer separation for a high
mobility two-dimensional electron gas sample.
The data in these figures are obtained from numerical evaluation of
Eq.~(\ref{tbe}) and Eq.~(\ref{tde}) with the input
parameters taken from the experiment of ref.~\cite{gramila1}.
In Fig.~(\ref{fb}) we see that $\tau_B^{-1}  \sim
T^2/d^4$ at low temperatures, as pointed out in
Ref.\cite{gramila1}. (The $d^{-4}$ dependence can be
recognized by noticing that the scattering rate decreases by
a factor of approximately four when the layer separation
increases by a factor of $\sqrt{2}$.) From Fig.~(\ref{fc}) one can see
that $\tau_{\Delta}^{-1}$ falls off more slowly
with both temperature and layer separation
as $T\rightarrow 0$.   The inset to Fig.~(\ref{fc})
establishes that for $T\ll T_{\tau}$,
$\tau_D^{-1} \sim - T^2{\rm ln}T/d^2$.  The dimensionless
temperature scale for Fig.~(\ref{fb}) is the Fermi
temperature ($T_F \equiv E_F / k_B$) which is about $60$K for
this sample, while
the dimensionless temperature scale for Fig.~(\ref{fc}) is
the disorder temperature
($ T_{\tau} \equiv \hbar / k_B \tau $) which is
$\sim 56$mK for this sample.   Comparing Fig.~(\ref{fb})
and Fig.~(\ref{fc}) we see that for this high-mobility
sample the disorder correction is smaller than
one part in $10^5$ at temperatures above $\sim 10mK$.

The origin of the temperature and layer dependence
seen in Fig.~(\ref{fb}) and Fig.~(\ref{fc}) can be
understood by looking at the limit of large layer separations where
the interlayer scattering rates can be evaluated analytically.
The evaluation of
$\tau_{LR}^{-1}$ at large layer separations and low
temperatures for the
disorder-free limit,
where the relation $\tau_B\propto T^2/d^4$ holds,
has been carried out previously
by several authors\cite{gramila1,solomon,jauho}.
(The results quoted in Ref.~(\cite{gramila1}) and
Ref.~(\cite{solomon}) are in error by a factor of two.)  We
rederive those results here to allow a comparison with the
disordered case.  For $T  \ll T_F$ and $ d \gg k_F^{-1}$
only the low frequency and long wavelength limit of
${\rm Im}\chi$ contributes importantly to Eq.~(\ref{r3}).
($k_F$ is the Fermi wavelength.) From Eq.~(\ref{xb}) it
follows that in this limit
$ {\rm Re} \chi (q,\omega) = dn/d\mu$, and ${\rm Im} \chi (q,\omega) =
dn/d\mu (2 \hbar  \omega /E_F) (k_F/q)$.  For $T \ll T_F$,  we can
ignore the contribution of ${\rm Im}\chi$ to screening the
interlayer interaction and it follows from Eq.~(\ref{rpa}) that
we can replace the interlayer interaction by
\begin{equation}
U_e(q) = { \pi e^2 q \over k_{TF}^2 \sinh{qd} }
\label{uesq}
\end{equation}
where $k_{TF} \equiv 2 \pi e^2  dn/d\mu$ is the single
layer Thomas-Fermi screening wavevector.  Note that for
$q \to 0$ the effective screening wavevector is $2 k_{TF}^2 d$
, which is proportional to the layer separation.  (For GaAs
$k_{TF} = 0.2 {\rm nm}^{-1}$ independent of electron density.)
With these approximations the integral over frequency and
wavevector are known and we obtain
\begin{equation}
\tau_B^{-1} = {- \pi\zeta(3)(k_BT)^2\over16\hbar\varepsilon_f
(k_{TF}d)^2(k_Fd)^2}.
\label{equ:tb2}
\end{equation}
In this result two powers of $d^{-1}$ may be associated with
the enhanced screening of the interlayer interaction at large
separations and two powers of $d^{-1}$ with the combination
of phase space considerations which cause the integrand to
to vary as $q^1$ for small $q$.
For layer separations larger than the mean free path the
electron response is diffusive over the entire range of
wavevectors contributing importantly to Eq.~(\ref{r3}).
For low temperatures it follows from Eq.~(\ref{xd}) that we
may use ${\rm Re} \chi (q,\omega) = dn / d \mu$ and
${\rm Im} \chi (q,\omega) = dn / d\mu [ \omega D q^2 /
( \omega^2 + (D q^2)^2) ]$.  Again we may ignore the
contribution to screening from ${\rm Im}\chi$ so that
the relevant limit of the screened interlayer interaction
is unchanged.  At small $\omega$, ${\rm Im}\chi \propto q^{-2}$
compared to the $q^{-1}$ of the ballistic case.  The
integrand of the wavevector integral thus goes as $q^{-1}$
at small wavevector.  This logarithmically divergent wavevector
integral is cutoff at $q \sim {\omega/D}^{1/2}$.
The remaining frequency integral is elementary and we obtain
\begin{equation}
\tau_D^{-1} = {- \pi(k_BT)^2{\rm ln}(T/T_{\tau})
\over 12 \hbar\varepsilon_f(q_{TF}d)^2(k_Fl)^2}.
\label{equ:td2}
\end{equation}
The change in the layer separation from $d^{-4}$ in the
ballistic case to $d^{-2}$ in the diffusive case can be traced
directly to the change in the wavevector dependence of
${\rm Im}\chi$ from $q^{-1}$ to $q^{-2}$.

For $d \ll l$ Eq.~(\ref{equ:td2}) correctly gives the
contribution to the drag from $q \ll l^{-1}$  and
Eq.~(\ref{equ:tb2}) gives the contribution from
$l^{-1} < q < d^{-1}$.  Because of the different
temperature dependence it is still true that the
contribution from the diffusive regime will dominate at
sufficiently low temperatures.  Comparing Eq.~(\ref{equ:td2})
and Eq.~(\ref{equ:tb2}) we can estimate the crossover temperature:
\begin{equation}
T_c \sim T_{\tau} \exp [ - 3 (l/d)^2 / 4 \zeta (3)]
\label{tcross}
\end{equation}
For high-mobility samples the diffusive enhancement of the
drag will be observable only at extremely low temperatures.
For example, in GaAs $T_{\tau}$ is about $0.2$K and $l \sim 10 \mu$
for a sample with a mobility of $\sim 10^6cm^{2}/sV$ and
a typical density.  For a layer separation of $\sim 500 A^{\circ}$
this implies that $T_c \sim 10^{-100} K$.
For the samples in the experiment of ref.~\cite{gramila1},
which are of extremely high mobility and small layer separations,
the value of $T_{\tau}$ is $56mK$ and the $d/l$ is about $10^{-3}$.
$ \tau_{\Delta}^{-1}$ is smaller than $\tau_B^{-1}$
by a factor of $10^{6\sim7}$ at $T\sim T_{\tau}$.
The correction term, $\tau_{\Delta}^{-1}$, will be
difficult to observe for accessible temperatures.
In samples with lower mobility and/or thicker
barriers between the layers, $\tau_{\Delta}^{-1}$ and $\tau_B^{-1}$
have comparable amplitudes for $T\sim T_{\tau}$ and the contribution
from the diffusive regime dominates at lower temperatures.
In Fig.~(\ref{fd}) we plot the relative contribution to
the drag from the diffusive regime vs. mobility for a layer
separation of $50 {\rm nm}$ and $n =1.5 \times 10^{11} {\rm cm}^{-2}$.
These results show that the effect of disorder
will become easily observable at typical low temperatures
for samples with mobilities below $\sim 10^{5} {\rm cm}^2 {\rm V}^{-1}
{\rm s}^{-1}$.

It is possible to fabricate double-layer systems in which
one layer is much more disordered than the other.
In particular one may have $l\gg d$ for one layer
and $l\le d$ for the other layer.
Following the same steps leading to Eq.~(\ref{equ:td2}),
it is possible to derive an expression for the low-temperature
transresistance in such a system
by using one diffusive response
function and one free electron response function:
\begin{equation}
\tau_{LR}^{-1} = {\pi^3(k_BT)^2\over72\hbar\varepsilon_f
(k_{TF}d)^2(k_Fd)(k_Fl)}.
\label{equ:tbd}
\end{equation}
The dependences on temperature and layer
separation, $\tau_{LR}^{-1} \sim
T^2/d^3$, are easily understood by comparing it to Eq.~(\ref{equ:tb2})
and Eq.~(\ref{equ:td2})
and noticing that the integrand in this case approaches a
constant at small wavevector transfers.
In Fig.~(\ref{fd}), results are shown for
the relative correction due to
disorder enhancement for the case where one layer
consists of free electrons while the other layer has
a finite mobility.
The relative correction is essentially independent of temperatures
at low temperatures and it is weaker than in
the case where both layers are disordered.

\section{summary}

Using the memory-function method,
we have derived an expression
for the transresistance of double layer systems
which is sufficiently general to treat the case of
disordered layers.  The expression has been evaluated
as a function of temperature, layer separation (d)
and in-plane mobility $\mu$.
Both the case where only one layer is disordered and
the case where both layers are disordered have been considered.
We find that the drag varies as $d^{-4}$, $d^{-3}$ and $d^{-2}$
at low temperatures for clean, single-layer disorder, and
double-layer disorder cases respectively.   In the case of
double-layer disorder the transresistance varies as
$ - T^2 \ln (T)$ at low temperatures, otherwise the
transresistance is proportional to $ T^2$.   The low-temperature
drag is proportional to $\mu^{-2}$ and $\mu^{-1}$ for
double-layer disorder and single-layer disorder respectively.
Except for the case of extremely low temperatures the
crossover from clean to disordered regimes occurs
when the mean-free-path within a layer becomes smaller
than the layer separation.
In very high mobility two-dimensional electron gas systems,
where the transresistance has been studied
experimentally up to the present,
the effect of disorder on the drag is negligible at available
temperatures.  We predict that the enhancement due to disorder
and the associated crossovers in the temperature-dependence
and layer-separation dependence will be observable at
low temperatures in moderate and low mobility samples.

\acknowledgments

The authors are grateful for stimulating
interactions with T.J. Gramila and J.P. Eisenstein and
with Jun Hu and Anthony Chan.  Informative
discussions with B.I. Altshuler during the initial stages of
this work and helpful communications with A.P. Jauho,
A.G. Rojo and G.D. Mahan are also acknowledged.
This work was supported by the National Science Foundation under
Grant No. DMR-9113911.

\newpage

\figure{Schematic diagrammatic rendering of the memory function
expression for the transresistivity to lowest order
in interlayer interactions.
(Eq.~(\ref{ffr}))
\label{fa}}

\figure{The interlayer scattering rate for a disorder-free
2DEG samples as a function of temperature.
$t \equiv T/T_{F}$ where $T_F$ is the Fermi temperature.
The solid curve was obtained for $d=315 A^{\circ}$
and the dashed curve was obtained for
$d=\sqrt{2}\times315A^{\circ}$.
These curves were calculated for typical and equal densities in the
two layers: $n_L=n_R=1.5\times10^{11}cm^{-2}V$ so that
$T_{F} \sim 60K$.
\label{fb}}

\figure{ The correction to the interlayer scattering rate
due to disorder-enhanced interactions at long wavelengths and
low temperatures.  Here $t=T/T_{\tau}$.
The insert shows that
$\tau_D^{-1} \propto T^2{\rm ln}T/d^2$ at low temperatures.
The solid curve is for $d=315 A^{\circ}$ and the dashed curve is for
$d=\sqrt{2}\times315A^{\circ}$.
Other parameters used are taken from the experiments
of Ref.(~\cite{gramila1}):
$n_L=n_R=1.5\times10^{11}cm^{-2}$, and $ \mu=3.4\times10^6cm^2/sV$.
The corresponding $T_{\tau}=56mK^{\circ}$ so that disorder
corrections would not be observable at available temperatures.
\label{fc}}

\figure{ Relative correction to the interlayer scattering
rate due to disorder-enhanced interactions for
several temperatures as a function of sample mobility.
These results were obtained for a typical layer separation:
$ d = 500A^{\circ}$.   The dashed line is for $T=1.0K$, the dotted
line for $T=0.1K$ and the solid line for $T=0.01K$.
The density
was taken from the experiments of Ref.(~\cite{gramila1}).
For these temperatures the diffusive contribution
become important for mobilities below about $ 10^{5}
{\rm cm}^2{\rm s}^{-1}{\rm V}^{-1}$.
The marked lines show the result for the case where one layer
has a finite mobility while the other layer has an infinitively
large mobility.
\label{fd}}

\end{document}